\documentclass[review]{elsarticle}
\usepackage{color}
\usepackage{amssymb}
\usepackage{lineno,hyperref}
\usepackage{float}
\modulolinenumbers[5]

\journal{Journal of \LaTeX\ Templates}

\newcommand{\Pom}{\mathbb{P}}                

\begin{document}

\begin{frontmatter}

\title{Predictions for future electron-ion colliders using the Balitsky--Kovchegov equation}


\author[CVUT]{Dagmar Bendova (\textit{dagmar.bendova@fjfi.cvut.cz})}
\address[CVUT]{Faculty of Nuclear Sciences and Physical Engineering,
Czech Technical University in Prague, Czech Republic}


\begin{abstract}
This contribution presents the latest predictions for several QCD processes at low-x in the color dipole picture which are of interest for current hadron-hadron and future electron-hadron colliders. The predictions are derived using the solution to the Balitsky-Kovchegov equation for proton and nuclear targets. Two different approaches to the nuclear case are studied: a solution obtained using a newly proposed type of initial condition which represents the initial state of a given nucleus and the solutions based on an initial condition representing a proton coupled to a Glauber-Gribov prescription. The influence from the different energy evolutions of these two approaches are studied in the following photo-nuclear processes: inclusive and diffractive DIS, coherent production of a J/psi meson in ultra-peripheral collisions, and the deeply virtual Compton scattering. By comparison to the available data from HERA and the LHC and to the other models inspired by the Color Glass Condensate framework, it is demonstrated that the future measurements will be useful to discriminate among different approaches to saturation physics.\newline
The contribution was presented at the Hot Quarks 2022 - Workshop for young scientists on the physics of ultrarelativistic nucleus-nucleus collisions, Dao House, Colorado, USA, October 11-17 2022.
\end{abstract}


\end{frontmatter}


\section{Introduction}
The deeply inelastic scattering (DIS) has been a very successful experimental tool to study the inner structure of hadrons, see e.g. recent HERA data from Ref.~\cite{H1:2018flt}. We therefore know that the proton is dominated by the valence quarks at low energies. However, with increasing energy, sea quarks and gluons emerge and contribute substantially to the proton structure; especially the gluon distribution grows rapidly with decreasing Bjorken $x$ and this growth is predicted to be tamed by their recombination in a phenomenon called parton saturation. Moreover, this structure is expected to be modified when the nucleons are bound inside a nucleus. Therefore, the exact behavior of the evolution of the hadronic structure at low $x$, its modification in nuclear matter, and other problems such as contribution from individual quarks and gluons to the overall mass and spin of hadrons still remain as open questions in QCD from both experimental and theoretical view; and so it is desired to invent QCD-based models in order to better understand QCD dynamics at high parton densities. This contribution therefore presents predictions for the measurements expected to be performed at future electron-ion colliders~\cite{Accardi:2012qut, AbelleiraFernandez:2012cc}, where such problems are going to be addressed.


\section{QCD phenomenology with the b-BK equation}

The DIS process in the regime of high energies can be viewed within the so called color dipole approach. In this model, the virtual photon, which mediates the interaction, is seen as splitting into the $q\bar{q}$ dipole (or a higher Fock state), which then strongly interacts with the target hadron via the gluon exchange. The cross section for the photon-hadron scattering can be then calculated as a convolution of the $\gamma \rightarrow q\bar{q}$ wave function and the dipole-hadron scattering amplitude, which encompasses the information about the strong interaction within the process, and therefore about the inner structure of the involved particles. The dipole amplitude can be either obtained from one of several QCD-based models on the market or as a solution to an evolution equation known as the Balitsky--Kovchegov (BK) equation \cite{Balitsky:1995ub,Kovchegov:1999ua}. Moreover, the dipole model can be extended to describe other QCD processes such as diffractive DIS and the production of exclusive states.

The BK equation describes a dressing of an initially bare color dipole with a cloud of gluons (these can be viewed as new dipoles in the limit of large $N_c$) and its form ensures a dynamical balance between the gluon emission and recombination towards the low-$x$ region. A necessary ingredient to solve this equation is the evolution kernel, which describes the probability of a gluon emission, and an appropriate initial condition. In the presented model, we used the collinearly-improved kernel, which imposes a time-ordering of the new emissions and suppresses unphysically large daughter dipoles, and a new form of initial condition introduced in Ref.~\cite{Cepila:2018faq}, which takes into account the end-points of the dipole and exponentially suppresses those far away from the target. Using these ingredients, the BK equation is solved with an explicit dependence on the impact parameter $b$ in Ref.~\cite{Cepila:2018faq,Bendova:2019psy}, the approach is denoted as b-BK. These solutions for the proton target suppress the Coulomb tails; an exponential growth of the dipole scattering amplitude towards large $b$, which in past prevented a direct phenomenological applications without any additional parameters. 

\begin{figure}[H]
\centering
\includegraphics[width=0.49\linewidth]{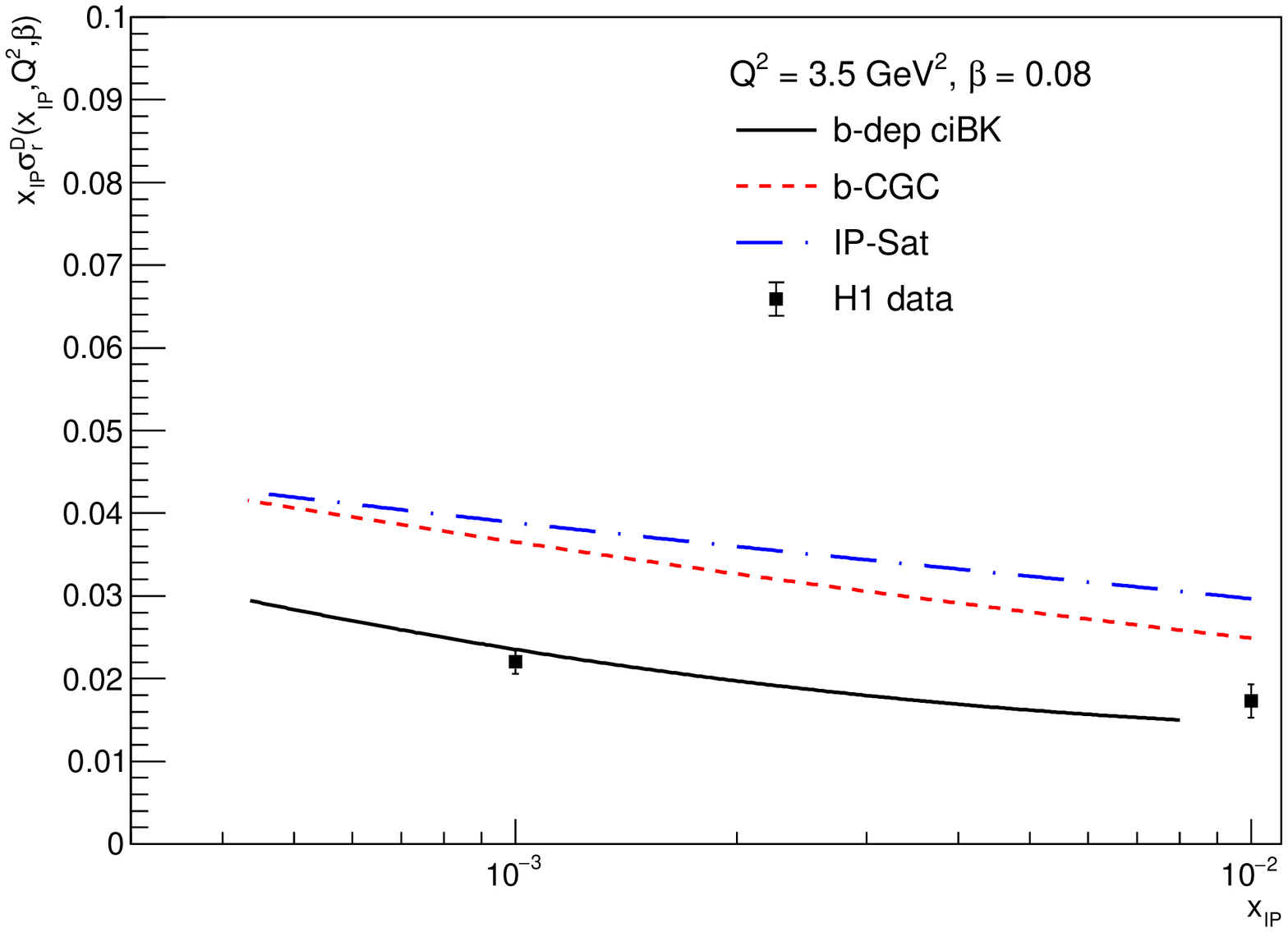}
\includegraphics[width=0.49\linewidth]{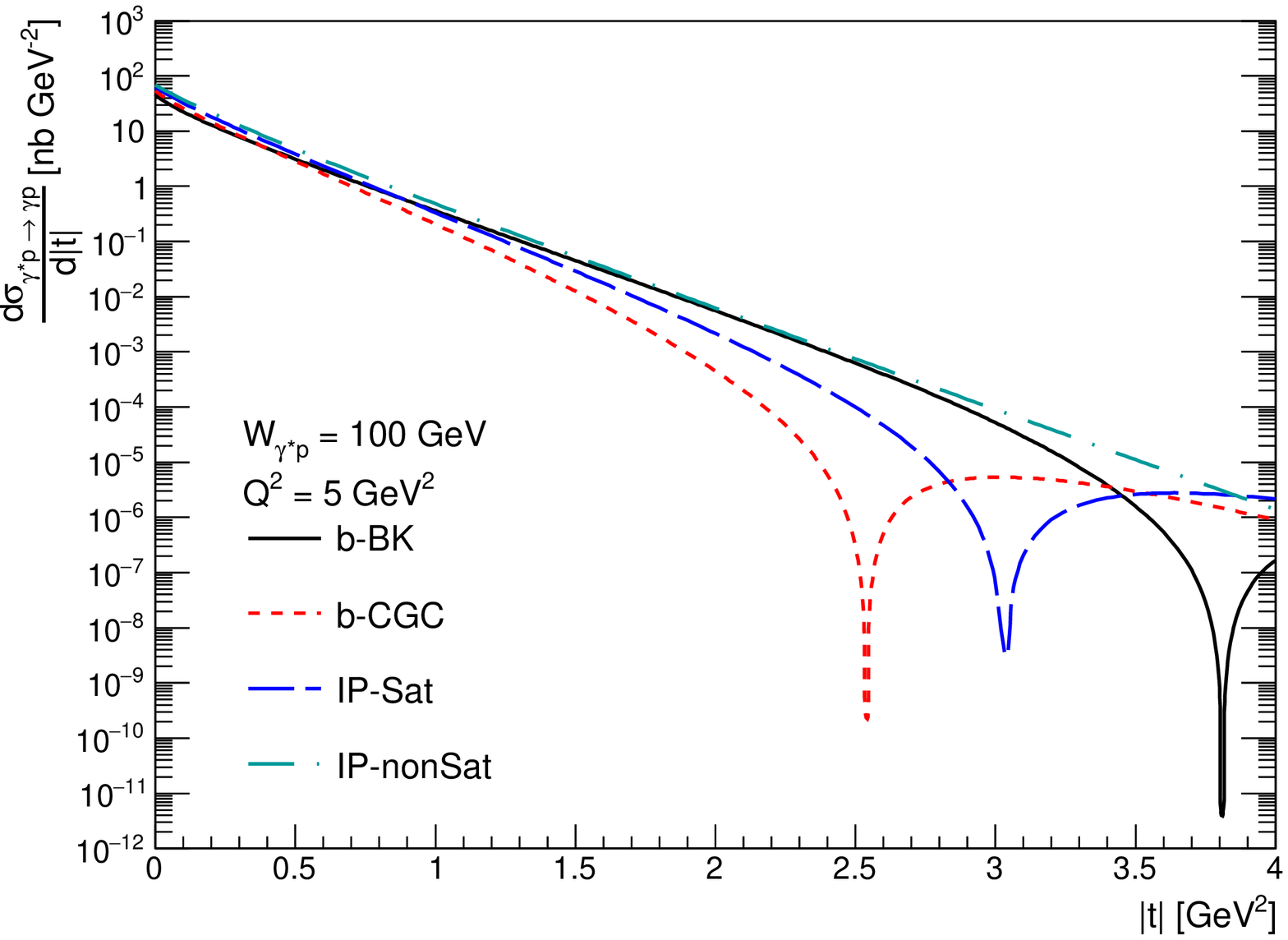}

\caption{(\textbf{a}) Predictions of the $x_{\Pom}$-dependence of the diffractive reduced cross section $\sigma_r^{D(3)}(x_{\Pom},\beta,Q^2)$ at compared with H1 data~\cite{Aaron:2012ad}. Figure from Ref.~\cite{Bendova:2020hkp}.
(\textbf{b}) Predictions for the $|t|$-distribution of the DVCS cross section in $ep$ scattering at $Q^2 = 5 \; \mathrm{GeV^2}$ using the b-BK and CGC-inspired models, for a CMS energy expected at the EIC. Figure from Ref.~\cite{Bendova:2022xhw}.
}
\label{fig:DDIS+DVCS}
\end{figure}

The resulting amplitudes are then applied to calculate various QCD processes. First, they were used to successfully describe a vast amount of data for DIS and production of vector mesons in $\gamma p$ interactions, see the plots in Ref.~\cite{Bendova:2019psy}. Later, they were used for comparison of our predictions to available data from HERA reported in Ref.~\cite{Aaron:2012ad} for diffractive reduced cross section; the agreement with the data is satisfactory as can be seen from Figure \ref{fig:DDIS+DVCS} (\textbf{a}) where b-BK model is in the full black line. We also provided predictions for the diffractive structure functions, all the results were published in Ref.~\cite{Bendova:2020hkp}. 

We also studied exclusive processes such as deeply virtual Compton scattering (DVCS). The results for this process were published in Ref.~\cite{Bendova:2022xhw}. An example of these results can be seen in Figure \ref{fig:DDIS+DVCS} (\textbf{b}) where the predictions are presented for the kinematic ranges expected to be measured at the EIC. These predictions are also compared to other phenomenological models and it is shown that the distinct models differ in their predictions. In case of predictions showed in Figure~\ref{fig:DDIS+DVCS} (\textbf{b}), the distinct models show a different position of the diffractive dips; a behavior which is expected to be measured in future experiments.

\begin{figure}[H]
\centering
\includegraphics[width=0.45\linewidth]{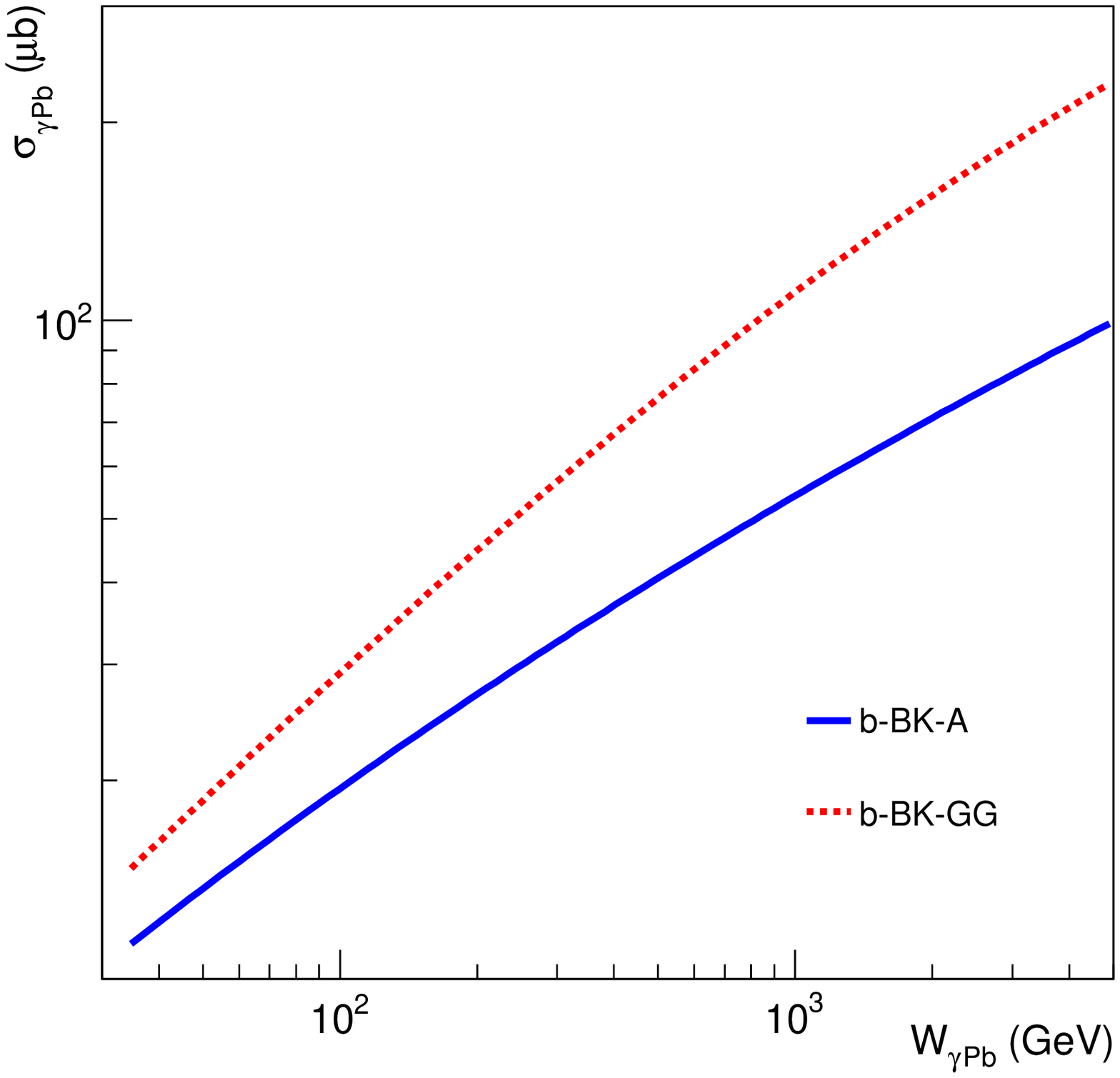}
\includegraphics[width=0.45\linewidth]{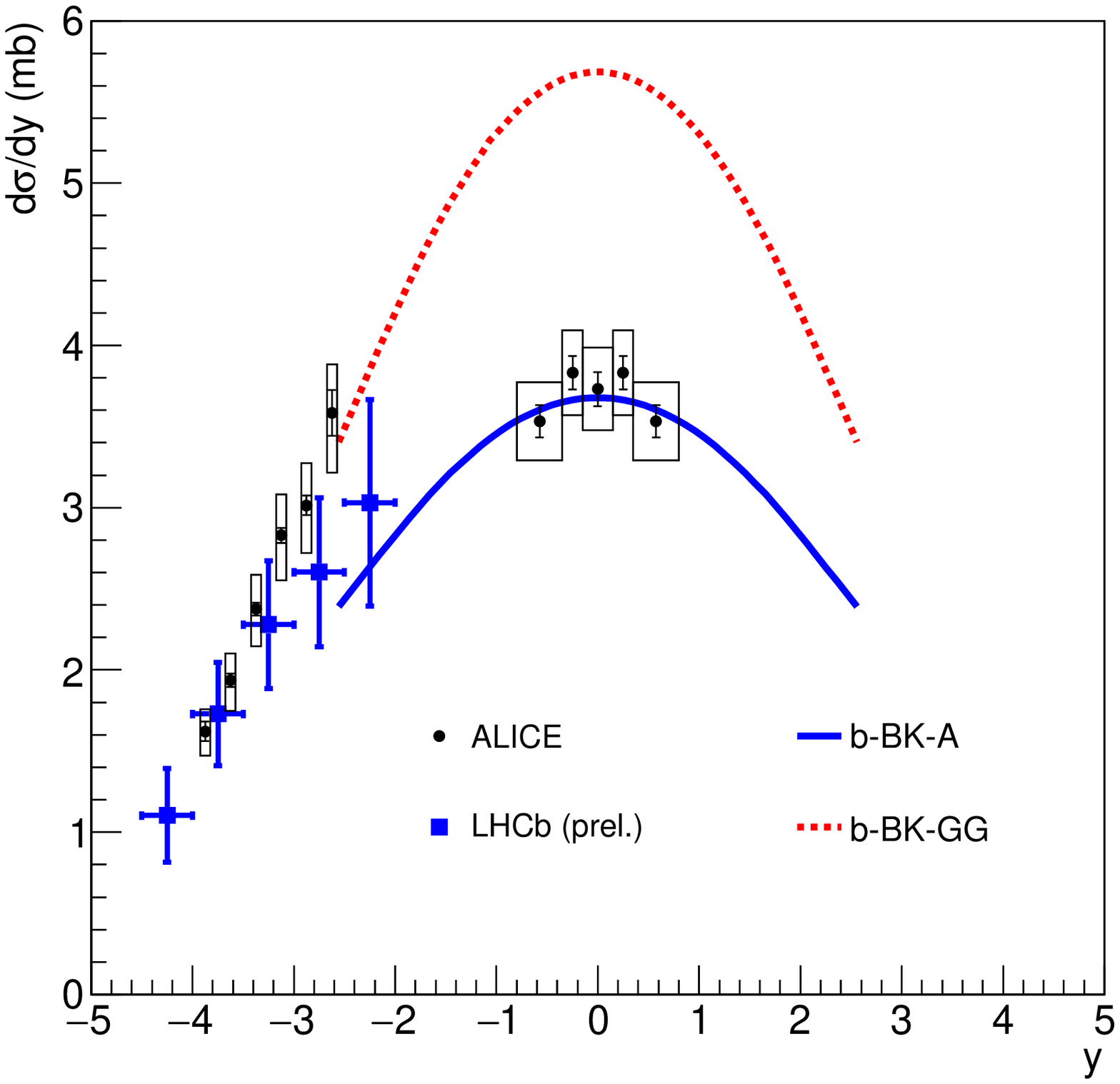}
\caption{(\textbf{a}) Predictions for the energy $W_{\gamma Pb}$-dependence of the coherent $J/\psi$ photoproduction off lead nuclei for the two nuclear BK models.
(\textbf{b}) Predictions for the rapidity dependence of the coherent $J/\psi$ photo-nuclear production, compared to LHC Run 2 data from Ref.~\cite{Acharya:2019vlb, Bursche:2018eni}. Figures from Ref.~\cite{Bendova:2020hbb}}.
\label{fig:Jpsi}
\end{figure}

In order to describe nuclear processes, two approaches to modify the calculation were proposed in Ref.~\cite{Cepila:2020xol}. First uses a solution of the b-BK equation for the proton target and couples it to a Glauber--Gribov approach (denoted as b-BK-GG) to obtain the dipole-nucleus scattering amplitude. The later evolves the nuclear dipole amplitude from an initial condition which represents a specific nucleus (denoted as b-BK-A). These two approaches were compared in several processes as in the proton case, e.g. diffractive DIS in Ref.~\cite{Bendova:2020hkp} and DVCS in Ref.~\cite{Bendova:2022xhw}, and were used to calculate predictions for future measurements at the EIC and LHeC. Moreover, Figure~\ref{fig:Jpsi} shows the predictions from Ref.~\cite{Bendova:2020hbb} for the coherent nuclear $J/\psi$ photoproduction for the two nuclear BK models. The panel (\textbf{a}) depicts the predictions for an energy $W_{\gamma Pb}$-dependence of the total $\gamma Pb \rightarrow J/\psi Pb$ cross section. It shows that the energy evolution within the two models is very different and moreover, the difference increases with energy, being approximately $30 \, \%$ at $W_{\gamma Pb} = 35 \; \mathrm{GeV}$ and reaching a factor of two at $W_{\gamma Pb} \sim \; \mathrm{TeV}$. The panel (\textbf{b}) shows the predictions for the rapidity-dependence of the coherent $J/\psi$ cross section compared to the LHC Run 2 data from Refs.~\cite{Acharya:2019vlb, Bursche:2018eni}. The same comparison for the Run 1 data (see Figure 2 in Ref.~\cite{Bendova:2020hbb}) suggested a preference for the b-BK-A approach, however the comparison to Run 2 data is rather inconclusive. Recently, ALICE experiment has published a new measurement of the coherent $J/\psi$ $|t|$-distribution photo-nuclear cross section, where the b-BK-A model shows a nice agreement with the measured data, see Figure 2 in Ref.~\cite{ALICE:2021tyx}. 

\section{Summary}
This contribution presented a sample of  latest predictions for various QCD processes within the color dipole approach. The results were obtained using the solutions to the impact-parameter dependent Balitsky-Kovchegov equation for proton and nuclear targets. The $b$-dependent dipole-proton and dipole-nuclear amplitudes were successfully applied into several QCD processes, namely the deeply inelastic scattering, diffractive DIS, production of vector mesons and the deeply virtual Compton scattering. These results are in reasonable agreement with HERA and LHC data, where applicable. Moreover, the predictions are of interest for future measurements at current (LHC) and planned (EIC and LHeC) facilities where these observables are expected to be measured with high precision. These measurements will therefore be able to study the hadronic structure at very low $x$ where non-linear effects are expected to manifest themselves. Thus will allow to distinguish among different approaches to model the saturation phenomenon and as a result, it will not only allow us to further constrain our models, but also to better understand the behavior of the structure of the proton and nuclei at high energies and the underlying QCD dynamics.

\vspace{6pt} 


\section*{Funding}
The work has been supported by the grant 22-27262S of the Czech Science Foundation (GACR).

\section*{Acknowledgements}
This contribution is papers listed in Refs.~\cite{Bendova:2019psy,Bendova:2020hkp,Bendova:2022xhw,Bendova:2020hbb} which were published in cooperation with J. Cepila, J. G. Contreras, V. Gonzalves, M. Matas, and C. R. Sena. I would like to thank the organizers of the Hot Quarks 2022 conference for giving me the opportunity to present these results.

\section*{Conflict of interest}
The author declares no conflict of interests.

\bibliography{HQ22_proc}

\begin{thebibliography}{10}
\expandafter\ifx\csname url\endcsname\relax
  \def\url#1{\texttt{#1}}\fi
\expandafter\ifx\csname urlprefix\endcsname\relax\def\urlprefix{URL }\fi
\expandafter\ifx\csname href\endcsname\relax
  \def\href#1#2{#2} \def\path#1{#1}\fi

\bibitem{H1:2018flt}
H.~Abramowicz, et~al., {Combination and QCD analysis of charm and beauty
  production cross-section measurements in deep inelastic $ep$ scattering at
  HERA}, Eur. Phys. J. C 78~(6) (2018) 473.
\newblock \href {http://arxiv.org/abs/1804.01019} {\path{arXiv:1804.01019}},
  \href {https://doi.org/10.1140/epjc/s10052-018-5848-3}
  {\path{doi:10.1140/epjc/s10052-018-5848-3}}.

\bibitem{Accardi:2012qut}
A.~Accardi, et~al., {Electron Ion Collider: The Next QCD Frontier}, Eur. Phys.
  J. A52~(9) (2016) 268.
\newblock \href {http://arxiv.org/abs/1212.1701} {\path{arXiv:1212.1701}},
  \href {https://doi.org/10.1140/epja/i2016-16268-9}
  {\path{doi:10.1140/epja/i2016-16268-9}}.

\bibitem{AbelleiraFernandez:2012cc}
J.~Abelleira~Fernandez, et~al., {A Large Hadron Electron Collider at CERN:
  Report on the Physics and Design Concepts for Machine and Detector}, J.Phys.
  G39 (2012) 075001.
\newblock \href {http://arxiv.org/abs/1206.2913} {\path{arXiv:1206.2913}},
  \href {https://doi.org/10.1088/0954-3899/39/7/075001}
  {\path{doi:10.1088/0954-3899/39/7/075001}}.

\bibitem{Balitsky:1995ub}
I.~Balitsky, {Operator expansion for high-energy scattering}, Nucl. Phys. B463
  (1996) 99--160.
\newblock \href {http://arxiv.org/abs/hep-ph/9509348}
  {\path{arXiv:hep-ph/9509348}}, \href
  {https://doi.org/10.1016/0550-3213(95)00638-9}
  {\path{doi:10.1016/0550-3213(95)00638-9}}.

\bibitem{Kovchegov:1999ua}
Y.~V. Kovchegov, {Unitarization of the BFKL pomeron on a nucleus}, Phys. Rev.
  D61 (2000) 074018.
\newblock \href {http://arxiv.org/abs/hep-ph/9905214}
  {\path{arXiv:hep-ph/9905214}}, \href
  {https://doi.org/10.1103/PhysRevD.61.074018}
  {\path{doi:10.1103/PhysRevD.61.074018}}.

\bibitem{Cepila:2018faq}
J.~Cepila, J.~G. Contreras, M.~Matas, {Collinearly improved kernel suppresses
  Coulomb tails in the impact-parameter dependent Balitsky-Kovchegov
  evolution}, Phys. Rev. D99~(5) (2019) 051502.
\newblock \href {http://arxiv.org/abs/1812.02548} {\path{arXiv:1812.02548}},
  \href {https://doi.org/10.1103/PhysRevD.99.051502}
  {\path{doi:10.1103/PhysRevD.99.051502}}.

\bibitem{Bendova:2019psy}
D.~Bendova, J.~Cepila, J.~Contreras, M.~Matas, {Solution to the
  Balitsky-Kovchegov equation with the collinearly improved kernel including
  impact-parameter dependence}, Phys. Rev. D 100~(5) (2019) 054015.
\newblock \href {http://arxiv.org/abs/1907.12123} {\path{arXiv:1907.12123}},
  \href {https://doi.org/10.1103/PhysRevD.100.054015}
  {\path{doi:10.1103/PhysRevD.100.054015}}.

\bibitem{Aaron:2012ad}
F.~Aaron, et~al., {Inclusive Measurement of Diffractive Deep-Inelastic
  Scattering at HERA}, Eur. Phys. J. C 72 (2012) 2074.
\newblock \href {http://arxiv.org/abs/1203.4495} {\path{arXiv:1203.4495}},
  \href {https://doi.org/10.1140/epjc/s10052-012-2074-2}
  {\path{doi:10.1140/epjc/s10052-012-2074-2}}.

\bibitem{Bendova:2020hkp}
D.~Bendova, J.~Cepila, J.~G. Contreras, t.~V.~P. Gon\c{c}alves, M.~Matas,
  {Diffractive deeply inelastic scattering in future electron-ion colliders},
  Eur. Phys. J. C 81~(3) (2021) 211.
\newblock \href {http://arxiv.org/abs/2009.14002} {\path{arXiv:2009.14002}},
  \href {https://doi.org/10.1140/epjc/s10052-021-09006-x}
  {\path{doi:10.1140/epjc/s10052-021-09006-x}}.

\bibitem{Bendova:2022xhw}
D.~Bendova, J.~Cepila, V.~P. Gon\c{c}alves, C.~R. Sena, {Deeply virtual Compton
  scattering at the EIC and LHeC: a comparison among saturation approaches},
  Eur. Phys. J. C 82~(2) (2022) 99.
\newblock \href {https://doi.org/10.1140/epjc/s10052-022-10059-9}
  {\path{doi:10.1140/epjc/s10052-022-10059-9}}.

\bibitem{Cepila:2020xol}
J.~Cepila, J.~Contreras, M.~Matas, {Predictions for nuclear structure functions
  from the impact-parameter dependent Balitsky-Kovchegov equation}, Phys. Rev.
  C 102~(4) (2020) 044318.
\newblock \href {http://arxiv.org/abs/2002.11056} {\path{arXiv:2002.11056}},
  \href {https://doi.org/10.1103/PhysRevC.102.044318}
  {\path{doi:10.1103/PhysRevC.102.044318}}.

\bibitem{Bendova:2020hbb}
D.~Bendova, J.~Cepila, J.~G. Contreras, M.~Matas, {Photonuclear $J/\psi$
  production at the LHC: Proton-based versus nuclear dipole scattering
  amplitudes}, Phys. Lett. B 817 (2021) 136306.
\newblock \href {http://arxiv.org/abs/2006.12980} {\path{arXiv:2006.12980}},
  \href {https://doi.org/10.1016/j.physletb.2021.136306}
  {\path{doi:10.1016/j.physletb.2021.136306}}.

\bibitem{Acharya:2019vlb}
S.~Acharya, et~al., {Coherent J/$\psi$ photoproduction at forward rapidity in
  ultra-peripheral Pb-Pb collisions at $\sqrt{s_{\rm{NN}}}=5.02$ TeV}, Phys.
  Lett. B 798 (2019) 134926.
\newblock \href {http://arxiv.org/abs/1904.06272} {\path{arXiv:1904.06272}},
  \href {https://doi.org/10.1016/j.physletb.2019.134926}
  {\path{doi:10.1016/j.physletb.2019.134926}}.

\bibitem{Bursche:2018eni}
A.~Bursche, {Study of coherent $J/\psi$ production in lead-lead collisions at
  $\sqrt{s_{\rm NN}} =5\ \rm{TeV}$ with the LHCb experiment}, Nucl. Phys. A 982
  (2019) 247--250.
\newblock \href {https://doi.org/10.1016/j.nuclphysa.2018.10.069}
  {\path{doi:10.1016/j.nuclphysa.2018.10.069}}.

\bibitem{ALICE:2021tyx}
S.~Acharya, et~al., {First measurement of the |$t$|-dependence of coherent
  $J/\psi$ photonuclear production}, Phys. Lett. B 817 (2021) 136280.
\newblock \href {http://arxiv.org/abs/2101.04623} {\path{arXiv:2101.04623}},
  \href {https://doi.org/10.1016/j.physletb.2021.136280}
  {\path{doi:10.1016/j.physletb.2021.136280}}.

\end{thebibliography}

\end{document}